\documentclass[12pt,preprint]{aastex}

 \newcommand{\text}     {}
 \newcommand{\Vec}   [1]{\vec #1}
 \newcommand{\cm}       {\rm cm}
 \newcommand{\erg}      {\rm erg}
 \newcommand{\Cs}       {C_{\rm s}}
 \newcommand{\AU}       {\rm AU}
 \newcommand{\eV}       {\rm eV}
 \newcommand{\G}        {\rm G}
 \newcommand{\kB}       {k_{\rm B}}
 \newcommand{\yr}       {\rm yr}

\shorttitle{Self-Sustained Ionization in Protoplanetary Disks}
\shortauthors{Inutsuka \& Sano}

\begin{document}

\title{Self-Sustained Ionization 
       and Vanishing Dead Zones\\
       in Protoplanetary Disks}

\author{ Shu-ichiro Inutsuka\altaffilmark{1} and 
         Takayoshi Sano\altaffilmark{2} } 

\altaffiltext{1}{Department of Physics, Kyoto University, 
                 Kyoto 606-8502, Japan;
                 inutsuka@tap.scphys.kyoto-u.ac.jp}
\altaffiltext{2}{Institute of Laser Engineering, Osaka University, 
                 Suita, Osaka 565-0871, Japan; 
                 sano@ile.osaka-u.ac.jp}

\begin{abstract}
We analyze the ionization state of the magnetohydrodynamically
 turbulent protoplanetary disks and
 propose a new mechanism of sustaining ionization.
First,
 we show that in the quasi-steady state of turbulence driven
 by magnetorotational instability in a typical protoplanetary disk 
 with dust grains, the amount of energy dissipation
 should be sufficient for providing the ionization energy 
 that is required
 for activating magnetorotational instability.
Second,
 we show that in the disk with dust grains
 the energetic electrons that compose electric currents 
 in weakly ionized gas can provide
 collisional ionization, depending on the actual saturation
 state of magnetorotational turbulence.
On the other hand, 
 we show that in the protoplanetary disks with the reduced effect
 of dust grains, the turbulent motion can homogenize the ionization
 degree, leading to the activation of magnetorotational instability 
 even in the absence of other ionization processes.
The results in this Letter indicate that most of the regions in
 protoplanetary disks remain magnetically active, 
 and we thus require
 a change in the theoretical modeling of planet formation.
\end{abstract}


\section{Introduction}
Protoplanetary disks are supposed to be the sites of 
 planet formation.  
In the standard models of planet formation, the effect of 
 the magnetic field is not seriously considered because 
 the ionization degree is supposed to be very small 
 at radii where the planet formation should take place. 
The low ionization is due to the high density of the midplane 
 of the disk, where the recombination rate is too high to 
 maintain ionization only through cosmic rays. 
However, if the ionization degree is sufficiently high,  
 the magnetorotational instability (MRI) is 
 important and the disk becomes turbulent  
 \citep{BalbusHawley1991,BalbusHawley1998,Stone et al. 2000}. 
The combination of Maxwell and Reynolds stresses
 in magnetohydrodynamic turbulence provides an efficient mechanism 
 for the radial transport of the angular momentum in the disk, 
 which results in gas accretion onto the central star through 
 the disk. 
Therefore, the determination of ionization degree 
 in realistic protoplanetary disks is very important 
 and remains the subject of extensive research   
 \citep{SanoMiyamaUmebayashiNakano2000,
        GlassgoldFeigelsonMontmerle2000, 
        FromangTerquemBalbus2002, 
        SalmeronWardle2003, 
        MatsumuraPudritz2003}. 
The existence of dust grains in the disk 
 has a drastic effect on the ionization degree, 
 because the recombination on the dust grains is 
 very efficient in the high-density regime. 
Most of the studies on the ionization degree with the effect of 
 dust grains show that there is a dead zone in which 
 gas effectively decouples with the magnetic field.  
The location of the dead zone spans from a few tenths of AU 
 to tens of AU from the central star 
 depending on the (column) density distribution of the disk.   

The existence of the dead zone 
 should have important implications for 
 the evolution of protoplanetary disks and 
 the formation of planets 
 (e.g., Gammie 1996, Matsumura \& Pudritz 2005)
 and their migration in the disk
 (e.g., Terquem 2003, Laughlin, Steinacker, \& Adams 2004, 
        Nelson \&Papaloizou 2004),  
 which requires more detailed analysis on the validity 
 of its existence and its precise location. 
In this Letter, we reanalyze the ionization degree 
 taking into account additional sources of ionization 
 that are not considered previously. 

\section{Ionization Required for MRI}
\paragraph{Ionization Degree} 
The linear stability analyses and nonlinear simulations 
 show that the MHD turbulence in the disk is driven by the MRI 
 if the magnetic Reynolds number is greater than about unity:   
$
	Re_{\rm M} = v_{\rm A}^2 /(\Omega \eta) \ga 1
$
 \citep{SanoInutsukaMiyama1998,FlemingStoneHawley2000,
        Sano & Stone 2002a, Sano & Stone 2002b}, 
 where $v_{\rm A}$ is the Alfv\'en velocity and 
 the resistivity $\eta$ can be expressed as a function of 
 the electron number density $x_e$ and temperature $T$: 
$
	\eta 
    \approx 2 \cdot 10^2 T^{1/2} / x_e . 
$
Thus, the condition is expressed as 
\begin{equation}
	Re_{\rm M} = \frac{v_{\rm A}^2}{\Omega \eta} 
	           \approx \frac{ \Cs H }{\beta \eta} 
               = \left( \frac{10}{\beta} \right) 
                 \left( \frac{x_e}{10^{-13}} \right)
                 \left( \frac{H}{0.1 \AU} \right) 
               \ga 1,  
                                              \label{eq:id}  
\end{equation}
 where $\Cs$ denotes the sound speed,   
 $H$ is the scale height of the disk, 
 and 
 $\beta=2\Cs^2/v_{\rm A}^2$ is the so called plasma $\beta$.  
In order to satisfy $Re_{\rm M} > 1$, 
 the number fraction of electrons should be larger than 
$
	x_e = 10^{-13}   
$
 at $r \approx 1$AU for typical protoplanetary disks with $\beta=10$. 
Numerical simulations on the saturation level of the nonlinear 
 development of MRI have shown that a typical value of 
 $\beta$ ranges from 10 to 100 
 (e.g., Hawley, Gammie, \& Balbus 1995, Sano et al. 2004). 
In this Letter  
 we adopt $\beta=10$ as a fiducial value  
 that corresponds to $B \approx 6$ Gauss for 
 $n=10^{15}\cm^{-3}$ and $T \approx 3 \cdot 10^2$ K.

\paragraph{Scaling Relation in High Density Regime with Dust grains}
%

\begin{figure}[hbt]
  \plotone{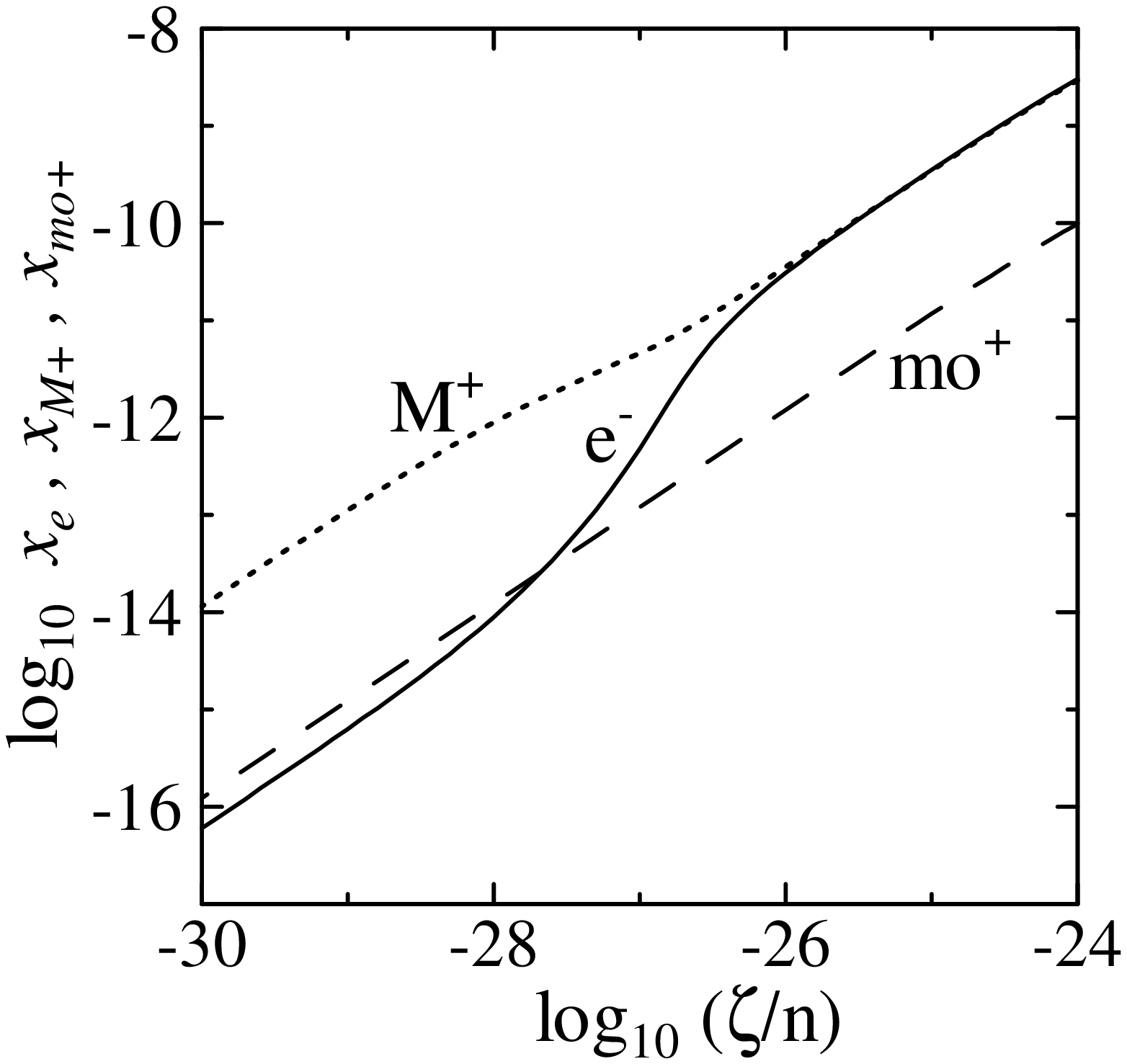}
  \caption{Fractional number densities of 
           electrons ($e^-$), metals (M$^+$), and molecules (mo$^+$) 
           as a function of 
           the ionization rate divided by the total number density  
           in the model with dust grains.  
           Recombinations on grain surfaces are taken into account 
           as well as the radiative and dissociative recombinations 
           in the gas phase.  
           The size of the dust grain is assumed to be 0.1 $\mu$m. 
           See Sano et al. (2000) for details.  
          }
          \label{fig:FracNumberDensity}
\end{figure}

%
%
The fractional number densities of various charged species 
 in the protoplanetary disks with dust grains 
 are essentially determined by a single parameter 
 that is the ionization rate $\zeta$
 divided by the number density $n \approx 2n_{\rm H_2}$   
 and calculated by Sano et al (2000). 
Figure \ref{fig:FracNumberDensity} plots the fractional number 
 densities of three important species 
 as a function of $\zeta/n$. 
The range of the fractional number density of electrons 
 required for MRI, 
 $x_e \approx 10^{-13}$, 
 corresponds to the normalized ionization rate 
$
    \zeta/n \approx 3 \cdot 10^{-28} \sec^{-1} \cm^3,  
$
 i.e.,   
\begin{equation}
  \zeta \approx 
  3 \cdot 10^{-13 }
    \left( \frac{  n  }{ 10^{15} \cm^{-3} } \right) \sec^{-1} . 
\end{equation}
This shows that for $ n =10^{15}~\cm^{-3}$ MRI requires 
 an ionization rate 30,000 times higher than 
 the standard cosmic-ray rate 
 ($\zeta_{\rm CR} \approx 10^{-17}$, see, e.g., Spitzer 1978). 
In the following, we consider whether this ionizing source is 
 available or not.  

\section{Feedback from MHD turbulence}
The column density of typical protoplanetary disks is so high that 
 the external ionizing radiation (UV and X-rays) cannot 
 penetrate deep into the midplane. 
We should consider the other mechanisms of ionization. 
\citet{SanoInutsuka2001} 
 have shown that the energy dissipation rate 
 $Q [\erg~\cm^{-3}\sec^{-1}]$
 in the turbulence driven by MRI in the disk can be calculated 
 by the time and spatial average of the $r\phi$-component of 
 the stress tensor: 
\begin{equation}
  Q = 
      \frac{{3\Omega }}{2}
      \left\langle {\left\langle  
        \rho v_R \delta v_\phi - \frac{{B_R B_\phi}}{{4\pi }}  
      \right\rangle } \right\rangle 
      \approx 0.03~\Omega~\langle\langle B^2 \rangle\rangle,  
\end{equation}
 where 
 the double angle brackets 
 denote time and spatial average  
 (see also Table 4 of Sano et al. 2004). 
%
%
Most of this energy dissipation is supposed to be used 
 for heating of the gas. 
In a quasi-steady state condition, 
 the input thermal energy is eventually converted 
 to radiation that escapes from the disk. 
Here we propose that some fraction of the energy dissipation 
 should be used for ionization. 

\begin{figure}[hbt]
  \plotone{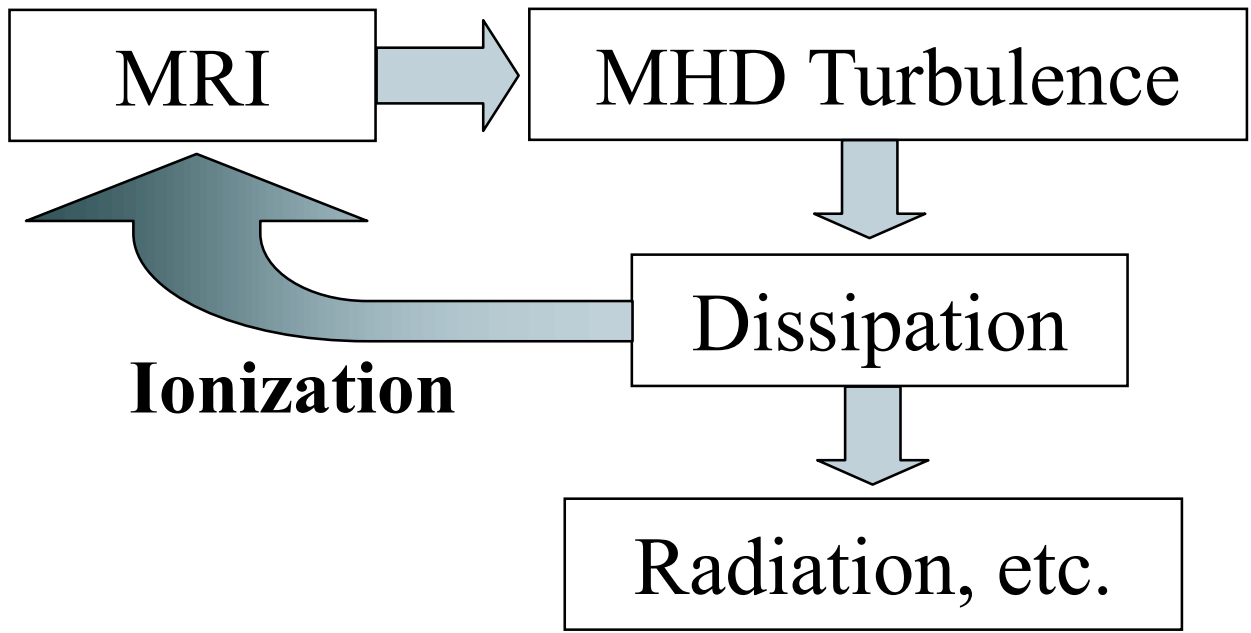}
  \caption{Schematic diagram for the feedback loop in 
           the turbulence driven by 
           magnetorotational instability (MRI). 
           Once the protoplanetary disk is sufficiently ionized, 
           the MRI occurs and  
           the magnetohydrodynamic turbulence develops and saturates. 
           At the saturated state, 
           most of the energy dissipation 
           results in the heating of the gas and its conversion 
           into the radiation that eventually escapes from the disk. 
           If only a small fraction of the energy dissipation 
           is used for ionization, the disk keeps this magnetic 
           activity (see eq.[\ref{eq:f_ionize}]). 
           One of the possible processes of ionization should be 
           the collision between the energetic electrons 
           and neutral particles (see Section 4). 
          }
          \label{fig:picture}
\end{figure}

\paragraph{Energy Budget for Sustaining Ionization}
First we consider whether the energy input available 
 in the MHD turbulence can provide sufficient energy for 
 the ionization required for MRI. 
The ratio $f_{\rm ionize}$ of the energy required for the ionization 
 to the energy available in the MRI-driven turbulence 
 can be calculated as  
\begin{equation}
 f_{\rm ionize} = \frac{\epsilon_{\rm ionize} \zeta n }{ Q }
  = 0.03
    \left( \frac{n             }{10^{15}cm^{-3}} \right)^2 
    \left( \frac{2\pi \yr^{-1} }{    \Omega    } \right)
    \left( \frac{     6 \G     }{      B       } \right)^2
    \left( \frac{\epsilon_{\rm ionize} }{ 13.6 \eV } \right) . 
                                                 \label{eq:f_ionize}
\end{equation}
This is sufficiently small for typical conditions 
 in protoplanetary disks even at 1 AU. 
Therefore, if the energy dissipation in MHD turbulence can be 
 used for ionization of the gas, 
 the positive feedback loop of MRI can maintain MHD turbulence in 
 the protoplanetary disks   
 as schematically shown in Figure \ref{fig:picture}.  
%
%

%
%

\section{Available Ionization Processes}
The next question is about the microscopic processes of ionization: 
 What is the process responsible for ionization in 
 weakly ionized dusty gas?  

\paragraph{Thermal Ionization}
Many numerical studies have shown that 
 the saturated state of MRI-driven turbulence is 
 the so-called high-$\beta$ plasma:  
 $\beta = 2\Cs^2/v_{\rm A}^2 \ga 10$;   
 i.e.,  the magnetic energy is smaller than 
        the thermal energy of the gas. 
Therefore, even a rapid conversion of all the magnetic energy 
 into thermal energy 
 (via, e.g., magnetic reconnection)  
 can raise the temperature of the gas only slightly.  
Thus, the magnetic dissipation does not seem to result in 
 thermal ionization.

Note, however, that this argument requires great care 
 because the thermal ionization of some alkali metals 
 takes place at quite low temperature, $T \approx 1000$K  
 (e.g., Umebayashi 1983). 
We also note that 
 the realistic saturation level of the magnetic field strength 
 is still unknown and remains to be determined theoretically. 
Obviously, this requires more detailed understanding of 
 the saturation mechanism of the MRI-driven turbulence, 
 which includes the understanding of the physics of rapid 
 magnetic reconnection in the case of a high-$\beta$ environment
 (e.g., Sano et al. 2004).

\paragraph{Electron Mobility} 
Let us assume that the thermal ionization is not promising, 
 and consider another mechanism. 
In weakly ionized plasma, electrons have a considerable 
 bulk velocity to maintain current and, hence, magnetic field. 
For sufficiently high density,  
 the collision frequency of electrons is larger than the 
 gyration frequency so that  
 the conductivity tensor is isotropic and diagonal. 
Thus, we can assume that the average velocity of the electrons 
 is essentially in the direction of the electric field 
 ($-\Vec{v}_e || \Vec{E}$)  
 in the comoving frame of the fluid. 
The averaged velocities of charged particles are related to the current 
 density $\Vec{j}$ via 
$
  \Vec{j} = \Sigma_i e q_i n_i \left< \Vec{v}_i \right> 
$
  or 
$ 
  e q_i n_i \left< v_i \right> \equiv f_i j , 
$
 where the angle brackets mean a phase-space average, 
 and $f_i$ is the fraction of the current due to 
 the $i$th species of charged particle. 
From one of Maxwell's equations, 
$
  4\pi \Vec{j}  = c \nabla \times \Vec{B} , 
$
  we estimate 
$ 
  j \approx cB/(2\lambda) , 
$
 where $\lambda$ is the typical length scale of the magnetic field. 
Thus, we can estimate the average velocities of the charged species: 
\begin{equation}
  | \left< v_e \right> | \approx  
    \frac{ cB }{ 2\lambda en_{\text{e}} } 
    =  
    42 \left( {\frac{B}{ 6 \G }} \right)
    \left( {\frac{0.03 {\rm AU} }{ \lambda }} \right)
    \left( {\frac{{10^{15} {\text{\cm}}^{ - {\text{3}}} }}
           {{n_{\text{H}} }}} \right)
       \left( \frac{ 10^{-13} }{ x_e } \right)
    {\rm km/s } , 
                                 \label{eq:vfromJ}
\end{equation}
 where we may assume $f_e \approx 1$ for 
 $\zeta/n \ga 3 \cdot 10^{-28} \sec^{-1} \cm^{3}$   
 and $\lambda \la 0.3H$ accords with the result 
 of numerical simulations in Sano et al. (2004). 
This average electron velocity $\left< v_e \right>$ 
 is surprisingly large. 
Next we show its implication for the distribution function 
 of electrons.

\paragraph{Energetic Electrons} 
To clarify the possible ionization process, 
 we should go into the details of microphysics. 
The electron distribution function $f(\Vec{p})$ 
 in weakly ionized plasma was already studied in detail  
 \citep{DruyvesteynPenning1940, LandauLifshitz1993}. 

If we choose the $z$-direction anti-parallel to $\Vec{E}$, 
 the electron distribution function is given by the following form 
 in spherical coordinates $(p,~\theta,~\phi)$ for the phase space: 
\begin{equation}
  f \approx f_0 (p) + f_1 (p)\cos (\theta ) , 
                                             \label{eq:fe}
\end{equation} 
\begin{equation}
  f_0 (p) = A \exp \left( { - \frac{3\epsilon^2}
                                   {\kB^2 T^2 \gamma ^2 } } 
                   \right),
\end{equation} 
\begin{equation}
  f_1 (p) = \frac{6 \epsilon}{\kB T \gamma}
            \sqrt{ \frac{m_e}{M} } 
            f_0 , 
\end{equation} 
where 
\begin{equation}
                  \epsilon = \frac{p^2}{2 m_e} ,
\end{equation}
\begin{equation}
  \gamma \equiv \frac{eEl}{\kB T} \sqrt{ \frac{M}{m_e} } \gg 1 , 
\end{equation} 
 and $m_e$ is the electron mass and 
 $M \approx 3670 m_e$ is the hydrogen molecule mass. 
The nomalization factor $A$ is propotional to the number density of 
 electrons. 
From these, we can calculate the average energy and the average 
 velocity in the direction of the electric field. 
\begin{equation}
  \left< \epsilon \right>  = 0.43 eEl\sqrt{ \frac{M}{m_e} } ,~~~
  \left< v_z \right>  = 0.9 \sqrt {\frac{{eEl}}
              {{m_e }}} \left( {\frac{{m_e }}
              {M}} \right)^{\frac{1}{4}}  
                                             \label{eq:e_e}
\end{equation} 
\begin{equation}
  \Rightarrow \left< \epsilon \right> 
      = 0.53 M \left< v_z \right>^2 . 
                                             \label{eq:mobil}
\end{equation} 
At first glance, it is surprising that the mass of the collision 
 partner $M$ appears in equation (\ref{eq:mobil}) 
 instead of the electron mass $m_e$. 
This corresponds to the very low mobility of electrons 
 in the collisional medium, which is due to $m_e \ll M$. 
If we insert the expression for 
 $\left< v_e \right> \approx 42 {\rm km}/\sec$ in equation 
 (\ref{eq:vfromJ}) into $\left< v_z \right>$
 in equation (\ref{eq:mobil}), we have 
 $\left< \epsilon \right> = 19 \eV$, 
 which is larger than the ionization potential of the hydrogen 
 molecule.  
Thus, most of the free electrons have sufficient energies 
 for ionization for this choice of parameters.  
This means that a significant fraction of the 
 collisions may result in ionization, 
 corresponding to a too high ionization rate. 
In reality, inelastic collisions with ionization/excitation losses 
 should modify the distribution function (eq.[\ref{eq:fe}]) 
 that is obtained by considering only elastic scattering. 
A higher ionization degree results in a lower average velocity of 
 electrons (see eq.[\ref{eq:vfromJ}]), 
 which in turn reduces the ionizing collision according to 
 equation (\ref{eq:mobil}). 
Therefore we expect that, in reality, 
 the ionization degree might be on the order of  
 $x_e \approx 10^{-13}$ .

\section{Yet Another Mechanism to Maintain Ionization Degree
         in Dust-Free Regions}
The analysis of the ionization degree in the previous section 
 is focused on the mid-plane of the disk, where external ionizing 
 radiation is not available 
 owing to the high column density of the disk.  
However, the gas density of the disk is a decreasing function 
 of height (distance from the mid-plane), and thus, 
 the surface layer of the disk is supposed to be 
 well ionized by the incident interstellar radiation field, 
 or by the radiation from the other stars in the cluster 
 environment (see, e.g., Adams \& Myers 2001). 
In the fully developed turbulence driven by MRI in the disk, 
 the fluid elements travel from place to place according to eddies.  
The inverse cascade of turbulent energy tends to 
 result in the state in which most of the power 
 is in the largest eddies 
 \citep{SanoInutsuka2001,SanoInutsukaTurnerStone2004}. 
This implies that each fluid element travels 
 from the surface layer of the disk to the mid-plane, 
 and vice versa. 
We should take into account this motion in the analysis 
 of the ionization degree. 
For simplicity, let us consider the region with a very 
 small amount of dust grains or the case in which the dust 
 grains have grown up to larger sizes so that the total 
 surface area of grains has become negligible. 
If we can ignore the recombination on the dust grains, 
 the recombination should happen in the gas phase. 
The rate equation for the fractional electron number 
 ($x_e=n_e/n$)
 is written as 
$
  dx_e/dt = \zeta - \sum_j \beta_j      x_e n_j
          = \zeta - \beta' n x_e^2 , 
$ 
 where 
 $\beta_j$ represents the coefficient 
 for the recombination of electrons with the $j$th species of 
 positively charged particles and   
$
 \beta' \equiv \sum_j \beta_j (x_j/x_e)
$
 is the effective recombination coefficient.  
Let us consider 
 the fluid element that travels from the surface 
 to the midplane. 
The approximate time evolution of the ionization degree 
 $x_e(t)$ can be obtained by solving the rate equation 
 with $\zeta=0$ and the approximately constant $\beta'$:  
\begin{equation}
  x_e(t)^{-1} = x_{e,0}^{-1} 
              + 10^{11} 
                \left( 
                       \frac{\beta'}{3 \cdot 10^{-12} \sec^{-1}\cm^3}
                \right)
                \left( 
                       \frac{ n }{ 10^{15}\cm^{-3} } 
                \right)
                \left( 
                       \frac{t}{1 \yr}   
                \right) ,
\end{equation}
 where $x_{e,0}$ is the initial ionization degree. 
For the high-density regime without the effect of grains, 
 the recombination with positively charged metals is 
 the most important process  
 \citep{SanoMiyamaUmebayashiNakano2000,
        FromangTerquemBalbus2002} .
That is, within an eddy turnover time ($\approx$ 1 yr at 1 AU), 
 the ionization degree keeps the level of $x_e \gg 10^{-13}$ 
 even without ionization due to cosmic rays,  
 if we adopt the standard value of 
 $\beta' \approx 3 \cdot 10^{-12} \cm^3 \sec^{-1}$ for metals
 and $x_{e,0} \gg 10^{-13}$.  
In other words, the ionized region penetrates into 
 the neutral region by the turbulent motions. 
In this way, the turbulent motions efficiently homogenize the 
 ionization degree if the recombination rate is sufficiently low. 
This spreading out of the weakly ionized region should be 
 an important mechanism for activating MRI in the region 
 that has initially low ionization. 
Note that this mechanism is not so effective in the region 
 with the standard interstellar dust grains, 
 because the recombination on the grain surface is 
 much faster than in the gas phase.

\section{Conclusions} 

Once the ionization degree becomes sufficiently high 
 (eq.[\ref{eq:id}]), 
 MRI-driven turbulence is the most promising mechanism 
 for gas accretion in protoplanetary disks. 
Conversely, once the MRI-driven turbulence becomes active, 
 it can consistently maintain the ionization degree without help of 
 other external ionizing sources:  
 the macroscopic energy dissipation in turbulence provides 
 sufficient energy for the ionization that is required for 
 activating MRI, 
 and microscopically the energetic electrons 
 in weakly ionized gas should provide 
 a sufficient rate of collisional ionization. 
The turbulent spreading of the ionized gas also provides 
 the mechanism of spreading 
 the magnetorotationally unstable region. 
Altogether we conclude that most of the region in the protoplanetary 
 disk should be sufficiently ionized, 
 thus, effectively removing the dead zone.  
The absence of the dead zone in the protoplanetary disks 
 has tremendous impact on the theoretical modeling of 
 planet formation. 

\acknowledgments
The authors thank the anonymous referee for the prompt and useful 
 comments. 
They also thank K. Shibata and T. Nakano for valuable discussions. 
The authors are supported by 
 the Grant-in-Aid (15740118, 16077202, 16740111, 17039005 ) 
 from the Ministry of Education, Culture, Sports, Science, 
 and Technology (MEXT) of Japan.
This work is supported by the Grant-in-Aid for the 21st Century COE 
 ``Center for Diversity and Universality in Physics'' from 
 MEXT of Japan.

\end{document}